\documentclass[lnbip]{svmultln}
\pdfoutput=1
\usepackage{makeidx}
\usepackage{epsfig}
\usepackage{amssymb}
\usepackage{amstext}
\usepackage{amsmath}
\usepackage{pslatex}
\usepackage{hyperref}

\usepackage{subfigure}
\usepackage{url}
\usepackage{graphicx}
\usepackage{color}
\usepackage{dsfont}
\usepackage{ifpdf}
\usepackage{color}
\newcommand{\daniel}[1]{\textcolor{red}{\textbf{#1}}}%
\newcommand{\kamel}[1]{\textcolor{blue}{\textbf{#1}}}%
\newcommand{\danielcut}[1]{}%
\newcommand{\kamelcut}[1]{}%

\graphicspath{{../op/}{../exp/}{../screenshots/}{../}}
\ifpdf
    \providecommand{\myfig}[1]{#1.pdf}
\else
   
   \providecommand{\myfig}[1]{#1.eps}
\fi

\begin{document}

\mainmatter              
\title{Web 2.0 OLAP: From Data Cubes to Tag Clouds}
\titlerunning{Web 2.0 OLAP}  
%
\author{Kamel Aouiche \and  Daniel Lemire \and Robert Godin}
\authorrunning{Kamel Aouiche et al.}   
%
\tocauthor{Daniel Lemire, Kamel Aouiche, Robert Godin}
\institute{Universit\'e du Qu\'ebec \`a Montr\'eal, 100 Sherbrooke West, Montreal, Canada,\\
\email{kamel.aouiche@gmail.com, lemire@acm.org, godin.robert@uqam.ca}}

\maketitle              

\begin{abstract}        
Increasingly, business projects are ephemeral. 
New Business Intelligence tools must support ad-lib data sources and
quick perusal.
Meanwhile, tag clouds are a popular community-driven visualization technique. 
Hence, we investigate tag-cloud views with support
for OLAP operations such as roll-ups, slices, dices, clustering, and drill-downs. As a case study, we
implemented an application where users can upload  data 
and immediately navigate through its ad hoc dimensions. 
To support social networking, views can be easily shared 
and embedded in other Web sites. 
Algorithmically, our tag-cloud views are approximate range top-k queries
over spontaneous data cubes. We present experimental evidence that 
iceberg cuboids provide adequate online approximations. 
We benchmark several browser-oblivious tag-cloud layout optimizations.
\end{abstract}

\section{Introduction}

 The Web~2.0, or Social Web, is about making available social software applications on the Web in an unrestricted manner. 
Enabling a wide range of distributed
individuals to collaborate on data analysis tasks 
may lead
to significant productivity gains~\cite{1240781,wattenberg2006dsd}.
Several companies, like  SocialText and IBM, 
are offering
Web~2.0  solutions dedicated to enterprise needs.
The data visualization Web sites Many~Eyes~\cite{manyeyes}
and Swivel~\cite{swivel} have become part of the Web~2.0 landscape: over 1~million data sets were uploaded to Swivel in less than 3~months~\cite{butler2007dsn}.

These Web~2.0 data visualization sites use traditional pie charts
and histograms, but also tag clouds. 
Tag clouds are a form of histogram 
which can represent the amplitude of over a hundred items by varying the font size. 
The use of hyperlinks makes tag clouds naturally
interactive. 
Tag clouds are used by many  Web~2.0 sites such as Flickr, del.icio.us and Technorati. Increasingly, e-Commerce sites
such as Amazon or O'Reilly Media, are using tag clouds to help their users navigate through aggregated data. 

\kamelcut{
\begin{figure}
\subfigure[Swivel]{
\includegraphics[width=0.45\columnwidth]{swivel}
}
\subfigure[Many Eyes]{
\includegraphics[width=0.45\columnwidth]{manyeyes}
}
\caption{\label{fig:web20tagclouds}Screen shots of popular Web~2.0 data sharing and visualization sites}
\end{figure}
}

Meanwhile, OLAP (On-Line Analytical Processing)~\cite{codd93} is a dominant paradigm
in Business Intelligence (BI).  OLAP allows domain experts to navigate
through aggregated data 
in a
multidimensional data model. Standard operations
include drill-down, roll-up, dice, and slice.
The data cube~\cite{graycube} model provides well-defined semantics and
performance optimization strategies.
However, OLAP requires much effort from database administrators even
after the data has been cleaned, tuned and loaded: schemas
must be designed in collaboration with users having fast
changing needs and requirements~\cite{583891,morzy2004qvm}. 
Vendors such as Spotfire, Business Objects and QlikTech have reacted
by proposing a new class of tools allowing end-user to customize their applications and to limit the need for centralized schema crafting~\cite{Havenstein2003}.


OLAP itself has never
been formally defined though rules have been proposed to recognize an OLAP
application~\cite{codd93}. In a similar manner, we propose  rules to recognize Web 2.0 OLAP
applications (see also Table~\ref{table:comparison}):
\begin{enumerate}
\item Data and schemas are provided autonomously by users. 
\item It is available as a Web application.
\item  It supports complete online interaction over
aggregated multidimensional data.
\item Users are encouraged to 
 collaborate.
\end{enumerate}


Tag clouds are  well suited for Web~2.0 OLAP.\@
They are flexible: a tag cloud can represent a dozen or hundred
different amplitudes.
And they are accessible: the only requirement is a browser
that can display different font sizes. They  also
 spark discussion~\cite{manyeyes2008}.

We describe a tag-cloud formalism,
as an instance of Web~2.0 OLAP.\@
Since we implemented a prototype, technical issues will be
discussed regarding application design.
In particular, we used iceberg cubes~\cite{253302} to  generate tag clouds 
online when the data and schema are provided extemporaneously.
Because tag clouds are meant
to convey a general impression,
presenting approximate measures and clustering is sufficient: we
propose specific metrics to measure the quality of tag-cloud approximations.
We conclude the paper with experimental results on real and synthetic data sets.

\begin{table}
\centering
\caption{\label{table:comparison}Conventional OLAP versus Web 2.0 OLAP}
\begin{scriptsize}
\begin{tabular}{|l|l|}
\hline
 Conventional OLAP & Web 2.0 OLAP \\
 \hline
 recurring needs &  ephemeral projects \\
 predefined schemas &   spontaneous schemas \\
 centralized design & user initiative \\
 histograms &  tag clouds \\
 plots and reports & iframes, wikis, blogs  \\
 access control & social networking\\
 \hline
\end{tabular}
\end{scriptsize}
\end{table}

\danielcut{
\begin{figure}
\centering\includegraphics[width=0.8\columnwidth]{generalmodel}
\caption{\label{fig:genmodel}Overview model of our Web 2.0 OLAP application}
\end{figure}
}

\section{Related Work}
 
There are decentralized models~\cite{1142476} and systems~\cite{1247631}
 to support collaborative data sharing
without a single schema.

According to Wu et al., it is difficult to navigate an OLAP schema without help; they have proposed a keyword-driven OLAP model~\cite{wu2007}.
There are several OLAP visualization techniques  including the
Cube Presentation Model (CPM)~\cite{maniatis2005pmn}, Multiple Correspondence
Analysis (MCA)~\cite{1150484} and other interactive systems~\cite{Techapichetvanich2005}.

Tag clouds 
 have been popularized
by the Web site Flickr launched in 2004. 
Several
optimization opportunities exist: similar tags can be
clustered together~\cite{kaser2007},
tags can be pruned automatically~\cite{hass:improving-tag-clouds}
or by user intervention~\cite{1124792},
tags can be indexed~\cite{1124792}, and so on. 
Tag clouds can be adapted to spatio-temporal data~\cite{1141859,1178692}.



\section{OLAP formalism}

\subsection{Conventional OLAP Formalism}
\begin{figure*}
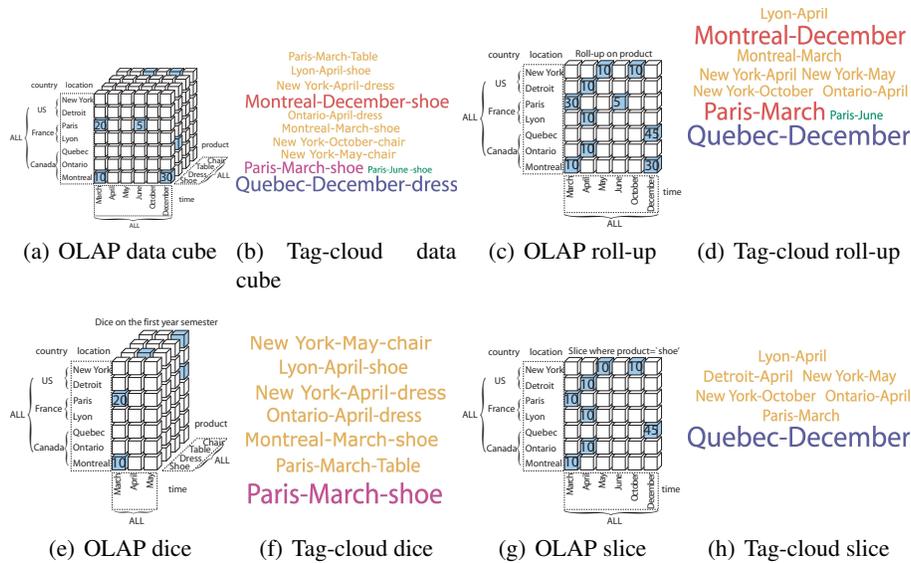

\begin{center}
  \subfigure[OLAP data cube]{\includegraphics[width=0.24\textwidth]{\myfig{olapcube}}\label{fig:OLAPCube}}
  \subfigure[Tag-cloud data cube]{\raisebox{0.5cm}{\includegraphics[width=0.24\textwidth]{\myfig{tccube}}}\label{fig:TCCube}}
  \subfigure[OLAP roll-up]{\includegraphics[width=0.24\textwidth]{\myfig{olaprollup}}\label{fig:OLAPRollup}}
  \subfigure[Tag-cloud roll-up]{\raisebox{1.1cm}{\includegraphics[width=0.24\textwidth]{\myfig{tcrollup}}}\label{fig:TCRollup}}
  \subfigure[OLAP dice]{\includegraphics[width=0.24\textwidth]{\myfig{olapdice}}\label{fig:OLAPDice}}
 \subfigure[Tag-cloud dice]{\raisebox{0.2cm}{\includegraphics[width=0.24\textwidth]{\myfig{tcdice}}}\label{fig:TCDice}}
  \subfigure[OLAP slice]{\includegraphics[width=0.24\textwidth]{\myfig{olapslice}}\label{fig:OLAPSlice}}
  \subfigure[Tag-cloud slice]{\raisebox{1cm}{\includegraphics[width=0.24\textwidth]{\myfig{tcslice}}\label{fig:TCSlice}}}
\end{center}
\caption{Conventional OLAP operations vs. tag-cloud OLAP operations}
\label{fig:expIceberg}
\end{figure*}

 Most OLAP engines rely on a data cube~\cite{graycube}. 
A data cube $\mathcal{C}$ contains a non empty set of $d$ dimensions
$\mathcal{D}=\{D_i\}_{1 \leq i \leq d}$ and a non empty set of measures
$\mathcal{M}$.
Data cubes are usually derived from a \emph{fact
table}  (see Table~\ref{tab:fact}) where each dimension and measure is a column and all rows (or facts) have
disjoint dimension tuples. 
Figure~\ref{fig:OLAPCube}
gives tridimensional representation of the data cube. 


\kamelcut{A data cube is made of a set of dimensions such as time, location,
product and so on. In most business applications, the number of dimensions is
relatively large (10 or more). Each dimension is made of a set of attribute
values: the location dimension might contain the list of the addresses of all
of the stores. These dimensions may have hierarchies. For example, the stores
might be grouped in cities and countries. The data cubes also contain measures
which are a special type of dimension whose values can be aggregated. Examples
of measures include cost, profit, value of sales, and so on. The data set is
usually represented as a fact table where each dimension and measure is a
column and all rows (or facts) have disjoint dimension tuples: a fact table is
commonly the result of a \textsc{group by} query (see Table~\ref{tab:fact}).}

\begin{table}\centering
\caption{Fact table example}
\begin{scriptsize}
\begin{tabular}{|llll|cc|}
\hline
\multicolumn{4}{|c|}{Dimensions} & \multicolumn{2}{|c|}{Measures}\\
\hline
location & time & salesman & product & cost & profit \\
\hline
Montreal & March & John & shoe & 100\$ & 10 \$ \\
Montreal & December & Smith & shoe & 150\$ & 30 \$ \\
Quebec & December & Smith & dress & 175\$ & 45 \$ \\
Ontario & April & Kate & dress & 90\$ & 10 \$ \\
Paris & March & John & shoe & 100\$ & 20 \$ \\
Paris & March & Marc & table & 120\$ & 10 \$ \\
Paris & June & Martin & shoe & 120\$ & 5 \$ \\
Lyon & April & Claude & dress & 90\$ & 10 \$ \\
New York & October & Joe & chair & 100\$ & 10 \$ \\
New York & May & Joe & chair & 90\$ & 10 \$ \\
Detroit & April & Jim & dress & 90\$ & 10 \$ \\
 \hline
\end{tabular}
\end{scriptsize}
\label{tab:fact}
\end{table}

\kamelcut{
\begin{figure}[H]\centering
 \includegraphics[width=1\columnwidth]{cube}
 \caption{Data cube example}
 \label{fig:cube}
 \end{figure}
}

Measures can be aggregated using several operators such as \textsc{average},
\textsc{max}, \textsc{min}, \textsc{sum}, and \textsc{count}. All of these measures and dimensions are
typically prespecified in a database schema. Database
administrators preaggregate views to accelerate queries.

The data cube supports the following operations: 
\begin{itemize}
\item A \emph{slice} specifies that you are only interested in some attribute
values of a given dimension. For example, one may want to focus on one specific
product (see Figure~\ref{fig:OLAPSlice}). 
Similarly, a \emph{dice} selects  ranges of attribute values (see Figure~\ref{fig:OLAPDice}).

\item A \emph{roll-up} aggregates the measures on coarser attribute values.
For example, from the sales given for every store, a user may want to see the
sales aggregated per country (see Figure~\ref{fig:OLAPRollup}). A \emph{drill-down} is
the reverse operation: from the sales per country, one may want to explore
the sales per store in one country. 


\end{itemize}

The various specific multidimensional views in Figure~\ref{fig:expIceberg}
are called \emph{cuboids}.

\subsection{Tag-Cloud OLAP Formalism}

A Web~2.0 OLAP application should be supported by a
flexible formalism that can adapt a wide range of data loaded
by users. 
Processing time must be reasonable and batch processing should be
avoided.




Unlike in conventional data cubes, we do not expect that most dimensions
have explicit hierarchies when they are loaded:
instead, users can  specify how the data is laid
out (see Section~\ref{tagclouddrawing}).
As a related issue,  the dimensions are not orthogonal
in general: there might be a ``City'' dimension as a
well as
``Climate Zone'' dimension. 
It is up to the user to 
organize the cities per climate zone or per country.



\begin{definition}[Tag] A tag is a term or phrase describing an object
with  corresponding non-negative weights determining its relative importance. Hence,
a tag is made of a triplet (term, object, weight).
\end{definition}

As an example, a picture may have been attributed the tags ``dog'' (12
times) and ``cat'' (20 times). In a Business Intelligence context, a tag may
describe the current state of a business. For example, the tags ``USA''
(16,000\$) and ``Canada'' (8,000\$)  describe the sales of a given product
by a given salesman. 

We can aggregate several attribute values, such as ``Canada'' and ``March,'' into a
single term, such as ``Canada--March.'' A tag composed of $k$~attribute values is called a
$k$-tag. 
Figure~\ref{fig:TCCube} shows a tag cloud representation of Table~\ref{tab:fact} using 3-tags.

Each tag $T$ is represented visually using
a font size, 
font color,
background color, area or motif, 
depending on its measure values.

\subsection{Tag-Cloud Operations}

\begin{figure}
\centering 
\includegraphics[width=1\columnwidth]{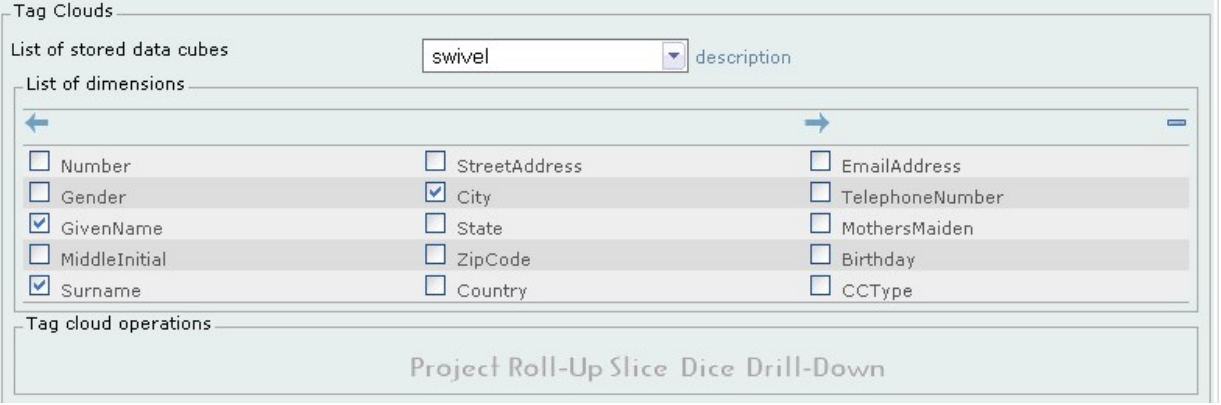}
\caption{User-driven schema design}
\label{fig:dimension_selection}
\end{figure}

In our system, users can upload data, select
a data set, and define a schema by choosing dimensions (see Figure~\ref{fig:dimension_selection}).
Then, users can apply various operations on the data using a menu bar. On the one hand, OLAP operations such as slice, dice, roll-up
and drill-down generate new tag clouds and new cuboids from existing
cuboids.
Figures~\ref{fig:TCRollup},~\ref{fig:TCDice}~and~\ref{fig:TCSlice}, show the results of a roll-up, a dice, and a slice
as tag clouds.
On the other hand, we can apply some operations on an existing tag cloud: sort by either the weights or the terms of tags, remove
some tags, remove lesser weighted tags, and so on.
We estimate that a tag 
cloud should not have  more than 150~tags.

\begin{figure}
\centering
\includegraphics[width=1\columnwidth]{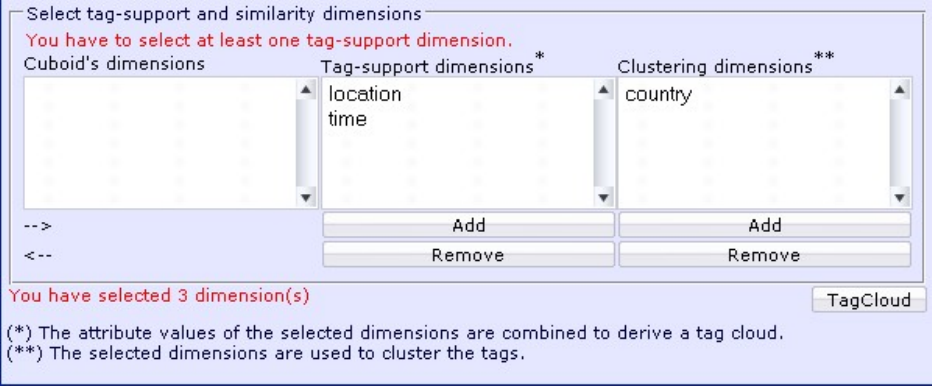}
\caption{Choosing similarity dimensions}
\label{fig:clustering}
\end{figure}

Tag-cloud layout has measurable
benefits when trying to convey a general impression~\cite{1240775}.
Hence, we wish to optimize the visual arrangement of tags.
Chen et al. propose the
computation of similarity measures between cuboids to help
users explore data~\cite{649755}: we apply this idea to define
similarities between tags. First of all, users are
asked to provide one or several dimensions they want to use
to cluster the tags. Choosing the ``Country'' dimension
would mean that the user wants the tags rearranged by
countries so that ``Montreal--April'' and ``Toronto--March'' are
nearby (see Figure~\ref{fig:clustering}).  The clustering dimensions selected by the user
together with the tag-cloud dimensions
form a cuboid: in our example, we have the dimensions ``Country,''
``City,'' and ``Time.''
Since a tag contains a set of attribute values, it has a corresponding \emph{subcuboid} defined
by  slicing the cuboid.

Several similarity measures can be applied
between subcuboids: Jaccard, 
Euclidean distance, cosine similarity, Tanimoto similarity,
Pearson correlation, Hamming distance, and
so on. Which similarity measure is best depends on the application at hand, so
advanced users should be given a choice.
Commonly, similarity measures take up
values in the interval $[-1,1]$. Similarity measures are expected to be
reflexive ($f(\texttt{a},\texttt{a})=1$), symmetric
($f(\texttt{a},\texttt{b})=f(\texttt{b},\texttt{a})$) and transitive: if
$\texttt{a}$ is similar to $\texttt{b}$, and $\texttt{b}$ is similar to
$\texttt{c}$, then $\texttt{a}$ is also similar to $\texttt{c}$.

\begin{figure}\centering
\includegraphics[width=0.6\columnwidth]{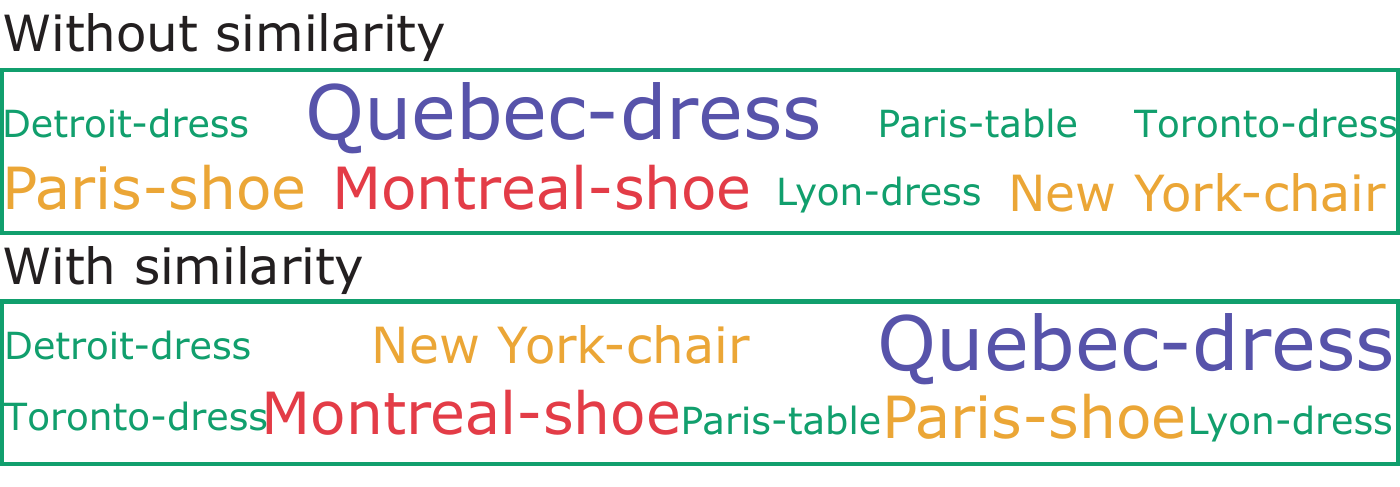}
\caption{Tag-cloud reordering based on similarity} \label{fig:tcsim}
\end{figure}

Recall that given two vectors $v$ and $w$, the cosine similarity measure
is defined as $\cos(v,w)=\sum_i v_i w_i / \sqrt{\sum_i v_i^2 \sum_i w_i^2}=
v/\vert v\vert \cdot w/ \vert w\vert$. The Tanimoto similarity is given by $\sum_i v_i
w_i / (\sum_i v_i^2 + \sum_i w_i^2 - \sum_i v_i w_i)$;
it becomes the
Jaccard similarity when the vectors have binary values. Both of these measures
are reflexive, symmetric and transitive. Specifically, the cosine similarity is
transitive by this inequality: $ \cos( v,z) \geq \cos(w, z)- \sqrt{1- \cos(v,w)^2}$.
To generalize the formulas from vectors to cuboids, it suffices
to replace the single summation by one summation per dimension.
Figure~\ref{fig:tcsim} shows an example of  tag-cloud reordering
to cluster similar tags. In this example, the ``City--Product'' tags
were compared according to the ``Country'' dimension. The
result is that the tags are clustered by countries.

\section{Fast Computation}
\label{sec:fastcomputation}
 
 Because only a moderate number of tags can be displayed, the computation of
tag clouds is a form of top-$k$ query: given any user-specified range of
cells, we seek the top-$k$ cells having the largest measures. There is a little
hope of answering such queries in near constant-time with respect to the number
of facts without an index or a buffer. Indeed, finding all
and only the elements with frequency exceeding a given frequency
threshold~\cite{corm:whats-hot} or merely finding the most frequent
element~\cite{237823} requires $\Omega (m)$~bits where $m$ is the number of
distinct items.

Various efficient techniques have been proposed for the
related range \textsc{max} problem~\cite{chazelle1988fad,Poon296}, but
they do not necessarily generalize.
Instead, for the range top-$k$ problem, 
we can  partition sparse data cubes into customized data structures to speed up
queries by  an order of
magnitude~\cite{luo2001rtb,loh2002amr,584806}. We can also answer range top-$k$ queries using
RD-trees~\cite{chung2007erm} or R-trees~\cite{seokjin:eer}.
In tag clouds, precision is not required and accuracy is less important;  only the most significant tags
are typically needed. Further, if all tags have similar weights, then any
subset of tag may form an acceptable tag cloud.



A strategy to speed up top-$k$ queries is to transform them  into
comparatively easier iceberg queries~\cite{253302}. For example, in computing
the top-10 ($k=10$) best vendors, one could start by finding all vendors with a
rating above 4/5. If there are at least 10~such vendors, then sorting this
smaller list is enough. If not, one can restart the query, seeking vendors with
a rating above 3/5. Given a histogram or selectivity estimates, we can reduce
the number of expected iceberg queries~\cite{donjerkovic1999pot}.
Unfortunately, this approach is not necessarily applicable to multidimensional
data since even computing iceberg aggregates once for each query may be
prohibitive. However, iceberg cuboids can still be put to good use. That is,
one materializes the iceberg of a cuboid, small enough to fit in main memory,
from which the tag clouds are computed. Intuitively, a cuboid representing the
largest measures is likely to provide reasonable tag clouds.
Users mostly notice tags with large font sizes~\cite{1240775}.
A good approximation
captures the tags having significantly larger weights.
To determine whether a tag cloud has such significant tags,
we can compute the  \emph{entropy}.


\begin{definition}[Entropy of a tag cloud]
Let $T \in \mathcal{T}$ be a tag from a tag cloud $\mathcal{T}$, then  $\textrm{entropy}(\mathcal{T})= -\sum_{T \in \mathcal{T}}  p(T) log(p(T))$
where $p(T)= \frac{\textrm{weight}(T)}{\sum_{x \in \mathcal{T}} \textrm{weight}(x)}$.
\end{definition}

\begin{figure}
\centering 
\includegraphics[width=1\columnwidth]{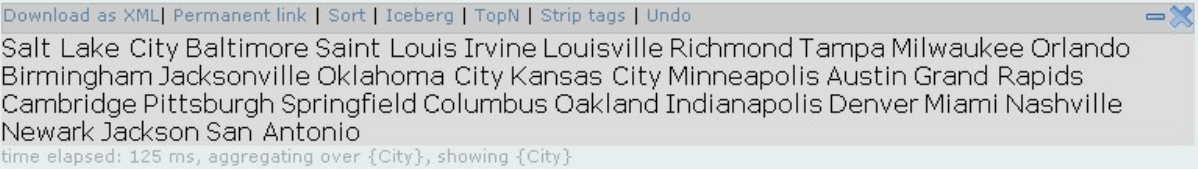}
\caption{Example of non informative tag cloud}
\label{fig:uniform}
\end{figure}

The entropy quantifies the disparity  of weights between tags. 
The \textbf{lower} the entropy, the \textbf{more} interesting the
corresponding tag cloud is. Indeed, tag clouds with uniform tag
weights have maximal entropy and are visually not very informative (see Figure~\ref{fig:uniform}). 

We can measure the quality of a low-entropy tag cloud by 
measuring false positives and negatives: false positive happens when a tag has been falsely added to a tag cloud whereas a false negative occurs when a tag is missing. These
measures of error assume that we limit the number of tags to a moderately
small number.
We use the following quality indexes; index values are in $[0,1]$ and a value of 0 is ideal; they are not applicable to high-entropy tag clouds.

\begin{definition} 
Given  approximate and exact tag clouds $A$ and $E$,
the false-positive and false-negative indexes are $\frac{\max_{t\in A, t\not \in E} \textrm{weight}(t)}{\max_{t\in A} \textrm{weight}(t)}$ and
 $\frac{\max_{t\in E, t\not \in A} \textrm{weight}(t)}{\max_{t\in E} \textrm{weight}(t)}$.
\end{definition}

\section{
Tag-Cloud Drawing}
\label{tagclouddrawing}
 While we can ensure some level of device-independent
displays on the Web, by using images or plugins, text display
in HTML may vary substantially from browser to another. There is no
common set of font browsers are required to support, and Web standards do
not dictate line-breaking algorithms or other typographical issues.
It is not practical to simulate the browser on a server.
Meanwhile, if we wish to remain accessible and to abide by open
standards, producing HTML and ECMAScript is the favorite option.

Given tag-cloud data, the tag-cloud drawing problem is
to optimally display the tags, generally using HTML, so that
some desirable properties are met, including the following:
(1) the screen space usage is minimized;
 (2) when applicable, similar tags are clustered together.
Typically, the width of the tag cloud is fixed, but its height
can vary.

For practical reasons, we do not wish for the server to send all of the data to
the browser, including a possibly large number of similarity measures between
tags. Hence,  some of the tag-cloud drawing computations must be
server-bound. There are  two possible architectures. The first scenario
is a browser-aware approach~\cite{kaser2007}:
 given the tag-cloud data provided
by the server, the browser sends back to the server some display-specific data,
such as the box dimensions of various tags using different font sizes. The
server then sends back an optimized tag cloud. The second approach is
browser-oblivious: the server optimizes the display of the tag cloud without
any knowledge of the browser by passing simple display hints. The browser can
then execute a final and inexpensive display optimization. While browser-oblivious
optimization is necessarily limited, it has reduced latency 
and it is easily cacheable. 

Browser-oblivious optimization can take many forms. For example, we could send
classes of tags and instruct the browser to display them on separate
lines~\cite{hass:improving-tag-clouds}. 
In our system, 
 tags are sent to the browser as an ordered list, using the
convention that successive tags are similar and should appear nearby. Given a similarity measure $w$ between tags, we want to minimize $\sum_{p,q}w(p,q) d(p,q)$
where  $d(p,q)$ is a distance function between the two tags in the list and the
sum is over all tags. Ideally, $d(p,q)$ should be the physical distance
between the tags as they appear in the browser; we model this distance with the index distance: 
if tag $a$ appears at index $i$ in the list and
tag $b$ appears at index $j$, their distance is the integer $\vert i - j\vert$.
This optimization problem is an instance of the NP-complete \textsc{minimum linear
arrangement} (MLA) problem: an optimal linear arrangement of a  graph $G=(V,E)$, is a
map $f$ from $V$ onto $\{1,2,\ldots,N\}$ minimizing $\sum_{u,v\in V} \vert
f(u)-f(v) \vert $.

\begin{proposition}
The browser-oblivious tag-cloud optimization problem is NP-Complete.
\end{proposition}

There is an
O($\sqrt{\log n} \log \log n$)-approximation for the MLA problem~\cite{1232252} in some instances. 
However, for our generic purposes, the greedy \textsc{Nearest Neighbor} (NN) algorithm
might suffice: insert any tag in an empty list, then repeatedly append a tag most similar to the  latest tag in the list,  until all tags have been inserted. 
It runs in O($n^2$) time where $n$ is the number of tags. 
Another heuristic for the MLA problem is the  \textsc{pairwise
exchange Monte Carlo} (PWMC) method~\cite{bhasker1987ola}: after applying NN,  you repeatedly consider the exchange of two tags chosen at random, permuting them if it reduces the MLA cost. Another \textsc{Monte Carlo} (MC) heuristic 
begins with the application of NN~\cite{johnson2004clb}: cut the list into two blocks at a random location, test if exchanging the two blocks reduces the MLA cost, if so proceed; repeat. 


Additional display hints can be inserted in this list. For example, if two tags
must absolutely be very close to each other, a \textsc{glued} token could be
inserted. Also, if two tags can be permuted freely in the list, then a
\textsc{permutable} token could be inserted: the list could take the form of a
PQ~tree~\cite{pqtrees}.

\kamelcut{
\begin{example}Given the tags A, B, C, D, E,  F, G, H, and the fact that
 A is similar to B and E, C is similar to D and E, we could output the following
 list: D, C, \textsc{glued}, E, A, B, F, \textsc{permutable}, G, \textsc{permutable}, H.
\end{example}
Once the client receives the ordered list, it can simply lay the tags out using
a greedy algorithm, doing its best to benefit from the display hints.
\daniel{What follows is insane. I tried implementing it and it is too difficult
in practice to be worth your time. Too much messing around with fragile
ECMAScript:} Laying out the tags can be done by alternating the order of the
tags, using a left-to-right layout for odd lines, and a right-to-left layout of
even lines, thus reducing the distance between tags that are consecutive in the
list. The last line can be either left aligned or right aligned depending on
whether there are an odd or even number of lines. When a \textsc{glue} has been
inserted, the browser may try to keep the two tags on the same line, whereas it
may try to permute tags when allowed to reduce the height of the tag cloud.

What kind of simple heuristic can we imagine to realize the embedding?
I'd like to benchmark browser-aware solutions against browser-oblivious
solutions. Not sure yet how to set it up.
I expect results to be much worst, but how much worse?

(Maybe for this paper, just do something silly, a prototypical heuristic and
see whether it makes sense.)

}

\section{Experiments}

 Throughout these experiments, we used the Java version 1.6.0\_02 from 
Sun Microsystems Inc. on an Apple MacPro machine with 2~Dual-Core Intel 
Xeon processors running at 2.66\,GHz and 2\,GiB of RAM.

\subsection{Iceberg-Based 
Computation}

\begin{figure}
\centering 
\includegraphics[width=0.6\columnwidth]{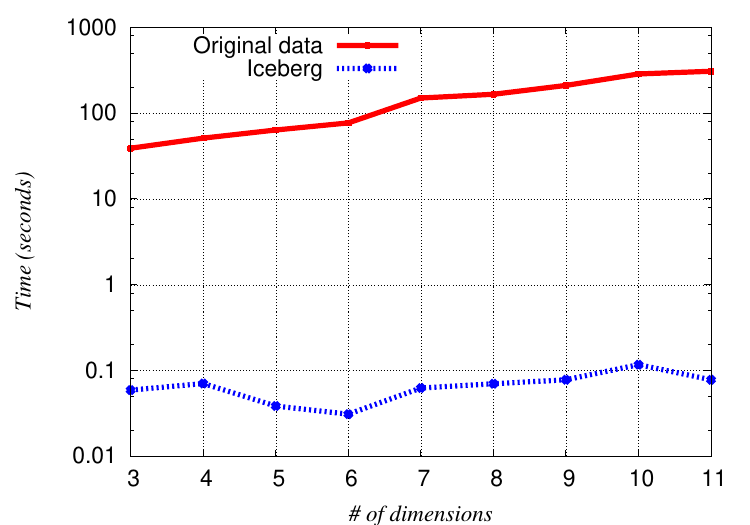}
\caption{Computing tag clouds from original data vs. icebergs: iceberg limit value set at 150 and tag-cloud size is 9 (US Income 2000).}
\label{fig:time}
\end{figure}

To validate the generation of tag clouds from icebergs, we have
run tests over the US~Income~2000 data set~\cite{KDDRepository} (42 dimensions and about $2 \times 10^5$ facts) as well as a
synthetic data set (18 dimensions and $2 \times 10^4$ facts)
provided by Swivel (\url{http://www.swivel.com/data_sets/show/1002247}).
Figure~\ref{fig:time} shows that while some tag-cloud computations
require several minutes, iceberg-based computations can
be much faster.

From each data set, we generated a 4-dimensional data cube.  
We used the COUNT function to aggregate data.
Tag clouds were computed from each data cube using the
iceberg approximation with different values of \emph{limit}: the number of
 facts
retained. We also implemented exact computations 
using temporary tables. We specified different values for tag-cloud size,
limiting the maximum number of tags. For each iceberg limit value
and tag-cloud size, we computed the entropy of the tag cloud, 
the false-positive and false-negative indexes, and
processing time for both of iceberg approximation and exact
computation.

\begin{figure*}
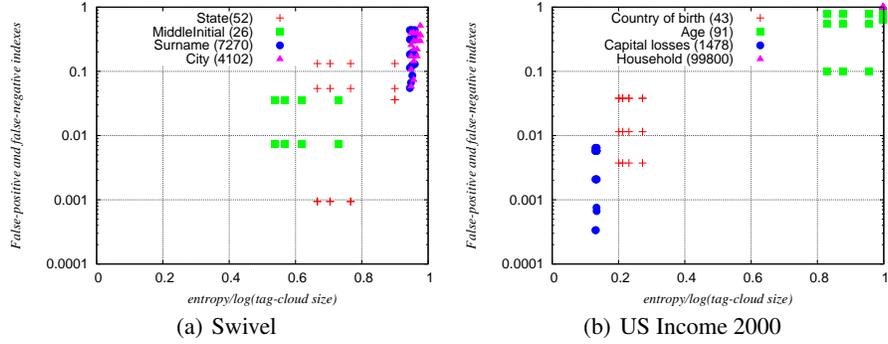

\centering
\subfigure[Swivel]{\includegraphics[width=0.49\textwidth]{\myfig{fnfp_vs_entropy_swivel}}\label{fig:expFNFNSwivel}}
\subfigure[US~Income~2000]{\includegraphics[width=0.49\textwidth]{\myfig{fnfp_vs_entropy_usincome}}\label{fig:expFNFPUsincome}}
\caption{False-negative and false-positive indexes (0 is best, 1 is worst), values under 0.0001 are not included} \label{fig:expIcebergIndexes}
\end{figure*}

We plotted in Figure~\ref{fig:expIcebergIndexes} the false-positive and false-negative indexes
as a function of the relative entropy (entropy/$\log(\textrm{tag-cloud size})$) using various iceberg
limit values (150, 600, 1200, 4800, and 19600) and various tag-cloud sizes (50, 100, 150, and 200), for a total
of 20~tag clouds per dimension.  The Y~axis is in a logarithmic scale. Points having their indexes equal to zero are not displayed.
As discussed in Section~\ref{sec:fastcomputation}, false-positive and false-negative indexes should be low when the entropy is low.
We verify that for low-entropy values ($<\frac{3}{4}\log(\textrm{tag-cloud size})$), the indexes are always close to zero which indicates a good 
approximation. 
Meanwhile, small iceberg cuboids can be processed
much faster. 

\begin{figure*}
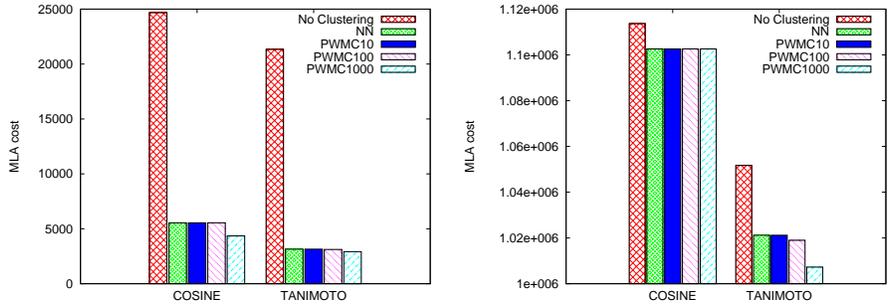

\centering
  \subfigure[Displaying dimension ``Givenname'' and clustering by ``State'' (Swivel)]{\includegraphics[width=0.49\textwidth]{\myfig{sim_swivel}}\label{fig:expSIMSwivel}}
  \subfigure[Displaying dimension ``HHDFMX'' and clustering by ``ARACE'' (US~Income~2000)]{\includegraphics[width=0.49\textwidth]{\myfig{sim_usincome}}\label{fig:expSIMUsincome}}
\caption{MLA costs for two examples: the PWMC heuristic was applied using 10, 100 and 1000 random exchanges.} \label{fig:expSIM}
\end{figure*}

\begin{figure*}
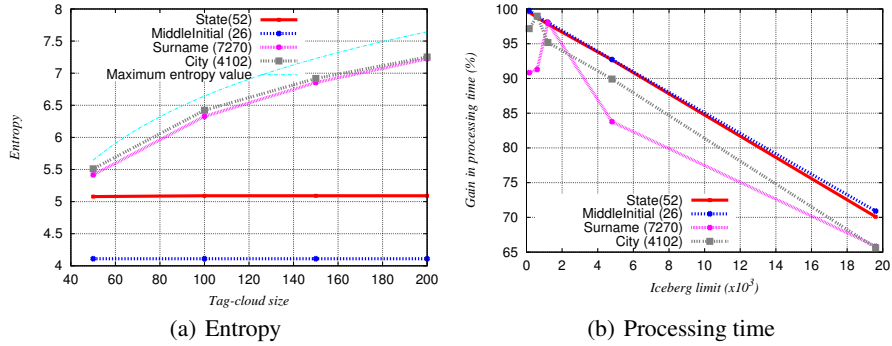

\begin{center}
  \subfigure[Entropy]{\includegraphics[width=0.49\textwidth]{\myfig{entropy_vs_tcs_limit1_swivel}}\label{fig:expEntropyVSTCSwivel}}
  \subfigure[Processing  time]{\includegraphics[width=0.49\textwidth]{\myfig{gain_vs_limit_TCS1_swivel}}\label{fig:expProcessingVSLimitSwivel}}
\end{center}
\caption{Benchmarking iceberg computation over Swivel}
\label{fig:expIcebergSwivel}
\end{figure*}
\begin{figure*}
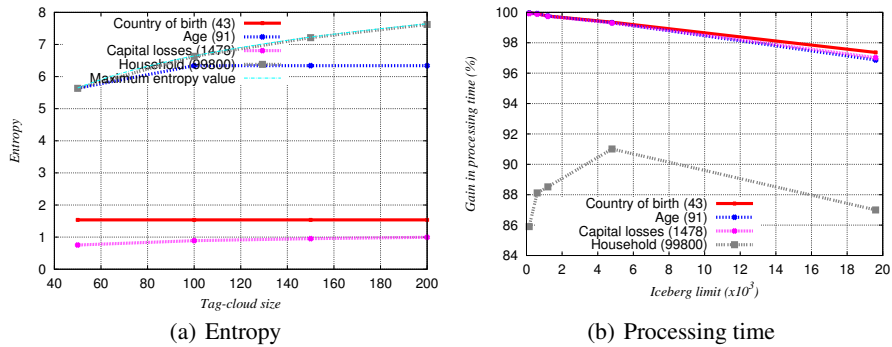

\begin{center}
  \subfigure[Entropy]{\includegraphics[width=0.49\textwidth]{\myfig{entropy_vs_tcs_limit1_usincome}}\label{fig:expEntropyVSTCUsincome}}
  \subfigure[Processing  time]{\includegraphics[width=0.49\textwidth]{\myfig{gain_vs_limit_TCS1_usincome}}\label{fig:expProcessingVSLimitUsincome}}
\end{center}
\caption{Benchmarking iceberg computation over US~Income~2000}
\label{fig:expUSincome}
\end{figure*}

Experimentally, we found that the entropy is not sensitive to the iceberg limit,
but it grows with the tag-cloud size (see 
Figures~\ref{fig:expEntropyVSTCSwivel}~and~\ref{fig:expEntropyVSTCUsincome}).
Naturally, the tag-cloud size is bounded by the cardinality of the chosen dimension. 

	We computed the relative gain in processing time due to the iceberg limit as $(t-t')/t$ where $t$
	is the time required for the exact computation whereas $t'$ is the time used by an iceberg computation (see
Figures~\ref{fig:expProcessingVSLimitSwivel}
and~\ref{fig:expProcessingVSLimitUsincome}). For these tests, the tag-cloud size was set to 150.
	Generally, the lower the iceberg limit value, the better the gain.
	High cardinality dimensions benefit less from a small iceberg limit.
	Also, the ratios of false positive and false negative decrease
	as the iceberg
limit increases. 
However, for low-cardinality dimensions, these ratios are often close to zero, so only
high-cardinality dimensions benefit from higher iceberg limits.
Hence,  you should choose an iceberg limit small or large depending
on whether you have a low or high cardinality dimension.

\subsection{Similarity Computation}
Using our two data sets, we tested the NN, PWMC, and MC 
heuristics using both the cosine and the Tanimoto similarity measures.
From data cubes made of all available dimensions, we 
used all possible 
1-tag clouds, using successively
all other dimensions as clustering dimension for a
total of $2 \times (18\times 17 + 42\times 41)=4056$~layout optimizations. The iceberg limit value was set at 150. 
The MC heuristic never fared better than NN, even when considering
a very large number of random block permutations: we rejected
this heuristic as ineffective.
However, as Figure~\ref{fig:expSIM} shows, 
the PWMC heuristic can sometimes significantly outperform NN
when a large number (1000)  of tag exchanges are considered,
but it only outperforms NN by more than 20\% in less 
than 5\% of all layout
optimizations. Meanwhile, table~\ref{table:similarities}
shows that 
if our objective is to reduce the MLA cost by
90\%, all heuristics are equivalent. However, it also
shows that 
PWMC can be several order of magnitudes 
slower than NN: NN is 10~times faster than PWMC with 100~exchanges and 
70~times faster than PWMC with 1000~exchanges.  
Computing the similarity function over an iceberg cuboid
was moderately expensive (0.07\,s) for a small iceberg
cuboid (limit set to 150~cells): the exact computation of the similarity
function can dwarf the cost of the heuristics (NN and PWMC)
over a moderately large data set. 
Informal tests suggest that NN computed over a small iceberg cuboid
provides significant visual layouts.


\begin{table}
\centering
\caption{\label{table:similarities}
Comparison of various \textsc{MLA} heuristics over the Swivel
data set using the cosine similarity measure (306 tag clouds). The 
running time is the average of 100~optimizations for tag clouds of size 150.
The number of tag clouds (out of 306) having at least a given gain is given.
}
\begin{tabular}{|l|c|ccc|}
\cline{2-5}
 \multicolumn{1}{l|}{}	& \multicolumn{1}{c}{NN}& \multicolumn{3}{|c|}{PWMC} \\
\cline{3-5}
  \multicolumn{1}{l|}{} &       & \multicolumn{1}{|c}{ 10}  & 100 & 1000  \\
 \hline
 time (s) & 0.003 & 0.01 & 0.03 & 0.2 \\
MLA gain $>0$\%   & 154   &  154  &   154  &   154  \\
MLA gain $>30$\%  & 143   &  143  &   145  &   148  \\
MLA gain $>70$\%  & 112   &  112  &   112  &   116  \\
MLA gain $>90$\%  &  97   &   97  &    97   &   99   \\
 \hline
\end{tabular}
\end{table}

\section{Conclusion 
}

 According to our experimental results, precomputing
a single iceberg cuboid per data cube allows to
generate adequate approximate tag clouds online.
Combined with modern Web technologies such as AJAX and
JSON, it provides a responsive application. However,
we plan
to make more precise the relationship between iceberg
cubes,  
entropy, dimension sizes, and our quality indexes.
Yet another approach to compute
tag clouds quickly may be to use a bitmap index~\cite{253268}. 
%
While we built a Web~2.0 with support for numerous
collaborations features such as permalinks, tag-cloud
embeddings with iframe elements, we still need to
experiment with live users.
Our approach to multidimensional tag clouds has been
to rely on $k$-tags. However, this approach might not
be appropriate when a dimension has a linear flow such
as time or latitude. A more appropriate approach is
to allow the use of a slider~\cite{1141859} tying
several tag clouds, each one corresponding to a given
attribute value. 

\subsubsection*{Acknowledgments}
 The second author is supported by NSERC grant~261437 and FQRNT grant~112381. 
The third author is supported by NSERC grant~OGP0009184 and FQRNT grant~PR-119731.
The authors wish to thank Owen Kaser from UNB for his contributions. 

\bibliographystyle{splncs}
{%
\bibliography{../../../bib/lemur}}

\begin{thebibliography}{10}

\bibitem{1240781}
Heer, J., Vi\'egas, F.B., Wattenberg, M.:
\newblock Voyagers and voyeurs: supporting asynchronous collaborative
  information visualization.
\newblock In: CHI '07. (2007)  1029--1038

\bibitem{wattenberg2006dsd}
Wattenberg, M., Kriss, J.:
\newblock Designing for social data analysis.
\newblock IEEE Transactions on Visualization and Computer Graphics
  \textbf{12}(4) (2006)  549--557

\bibitem{manyeyes}
{IBM}:
\newblock {Many Eyes}.
\newblock \url{http://services.alphaworks.ibm.com/manyeyes/} (2007) [Online;
  accessed 7-6-2007].

\bibitem{swivel}
{Swivel, Inc}:
\newblock Swivel.
\newblock \url{http://www.swivel.com} (2007) [Online; accessed 7-6-2007].

\bibitem{butler2007dsn}
Butler, D.:
\newblock Data sharing: the next generation.
\newblock Nature \textbf{446}(7131) (2007)  1--10

\bibitem{codd93}
Codd, E.F.:
\newblock Providing {OLAP} (on-line analytical processing) to user-analysis: an
  {IT} mandate.
\newblock Technical report, E. F. Codd and Associates (1993)

\bibitem{graycube}
Gray, J., Bosworth, A., Layman, A., Pirahesh, H.:
\newblock Data cube: A relational aggregation operator generalizing group-by,
  cross-tab, and sub-total.
\newblock In: ICDE '96. (1996)  152--159

\bibitem{583891}
Body, M., Miquel, M., B\'edard, Y., Tchounikine, A.:
\newblock A multidimensional and multiversion structure for {OLAP}
  applications.
\newblock In: DOLAP '02. (2002)  1--6

\bibitem{morzy2004qvm}
Morzy, T., Wrembel, R.:
\newblock On querying versions of multiversion data warehouse.
\newblock In: DOLAP '04. (2004)  92--101

\bibitem{Havenstein2003}
Havenstein, H.:
\newblock {BI} vendors seek to tap end-user power: New class of tools built to
  reap user knowledge for customizing analytic applications.
\newblock InfoWorld \textbf{22} (2003)  20--21

\bibitem{manyeyes2008}
Vi\'egas, F., Wattenberg, M., McKeon, M., van Ham, F., , Kriss, J.:
\newblock Harry potter and the meat-filled freezer: A case study of spontaneous
  usage of visualization tools.
\newblock In: HICSS 2008. (2008)  1--10

\bibitem{253302}
Carey, M.J., Kossmann, D.:
\newblock On saying ``enough already!'' in {SQL}.
\newblock In: SIGMOD'97. (1997)  219--230

\bibitem{1142476}
Taylor, N.E., Ives, Z.G.:
\newblock Reconciling while tolerating disagreement in collaborative data
  sharing.
\newblock In: SIGMOD '06, New York, NY, USA, ACM (2006)  13--24

\bibitem{1247631}
Green, T.J., Karvounarakis, G., Taylor, N.E., Biton, O., Ives, Z.G., Tannen,
  V.:
\newblock {ORCHESTRA}: facilitating collaborative data sharing.
\newblock In: SIGMOD '07, New York, NY, USA, ACM (2007)  1131--1133

\bibitem{wu2007}
Wu, P., Sismanis, Y., Reinwald, B.:
\newblock Towards keyword-driven analytical processing.
\newblock In: SIGMOD '07. (2007)  617--628

\bibitem{maniatis2005pmn}
Maniatis, A., Vassiliadis, P., Skiadopoulos, S., Vassiliou, Y., Mavrogonatos,
  G., Michalarias, I.:
\newblock A presentation model \& non-traditional visualization for {OLAP}.
\newblock International Journal of Data Warehousing and Mining \textbf{1}
  (2005)  1--36

\bibitem{1150484}
{Ben Messaoud}, R., Boussaid, O., {Loudcher Rabas{\'e}da}, S.:
\newblock Efficient multidimensional data representations based on multiple
  correspondence analysis.
\newblock In: KDD'06. (2006)  662--667

\bibitem{Techapichetvanich2005}
Techapichetvanich, K., Datta, A.:
\newblock Interactive visualization for {OLAP}.
\newblock In: ICCSA '05. (2005)  206--214

\bibitem{kaser2007}
Kaser, O., Lemire, D.:
\newblock Tag-cloud drawing: Algorithms for cloud visualization.
\newblock In: WWW 2007 -- Tagging and Metadata for Social Information
  Organization. (2007)

\bibitem{hass:improving-tag-clouds}
{Hassan-Montero}, Y., {Herrero-Solana}, V.:
\newblock Improving tag-clouds as visual information retrieval interfaces.
\newblock In: InSciT'06. (2006)

\bibitem{1124792}
Millen, D.R., Feinberg, J., Kerr, B.:
\newblock Dogear: Social bookmarking in the enterprise.
\newblock In: CHI '06. (2006)  111--120

\bibitem{1141859}
Russell, T.:
\newblock cloudalicious: folksonomy over time.
\newblock In: JCDL'06. (2006)  364--364

\bibitem{1178692}
Jaffe, A., Naaman, M., Tassa, T., Davis, M.:
\newblock Generating summaries and visualization for large collections of
  geo-referenced photographs.
\newblock In: MIR '06. (2006)  89--98

\bibitem{1240775}
Rivadeneira, A.W., Gruen, D.M., Muller, M.J., Millen, D.R.:
\newblock Getting our head in the clouds: toward evaluation studies of
  tagclouds.
\newblock In: CHI'07. (2007)  995--998

\bibitem{649755}
Chen, Q., Dayal, U., Hsu, M.:
\newblock {OLAP}-based data mining for business intelligence applications in
  telecommunications and e-commerce.
\newblock In: DNIS '00. (2000)  1--19

\bibitem{corm:whats-hot}
Cormode, G., Muthukrishnan, S.:
\newblock What's hot and what's not: tracking most frequent items dynamically.
\newblock ACM Trans. Database Syst. \textbf{30}(1) (2005)  249--278

\bibitem{237823}
Alon, N., Matias, Y., Szegedy, M.:
\newblock The space complexity of approximating the frequency moments.
\newblock In: STOC '96. (1996)  20--29

\bibitem{chazelle1988fad}
Chazelle, B.:
\newblock A functional approach to data structures and its use in
  multidimensional searching.
\newblock SIAM J. Comput. \textbf{17}(3) (1988)  427--462

\bibitem{Poon296}
Poon, C.K.:
\newblock Dynamic orthogonal range queries in {OLAP}.
\newblock Theoretical Computer Science \textbf{296}(3) (2003)  487--510

\bibitem{luo2001rtb}
Luo, Z.W., Ling, T.W., Ang, C.H., Lee, S.Y., Cui, B.:
\newblock Range top/bottom k queries in {OLAP} sparse data cubes.
\newblock In: DEXA'01. (2001)  678--687

\bibitem{loh2002amr}
Loh, Z.X., Ling, T.W., Ang, C.H., Lee, S.Y.:
\newblock Adaptive method for range top-k queries in {OLAP} data cubes.
\newblock In: DEXA'02. (2002)  648--657

\bibitem{584806}
Loh, Z.X., Ling, T.W., Ang, C.H., Lee, S.Y.:
\newblock Analysis of pre-computed partition top method for range top-k queries
  in {OLAP} data cubes.
\newblock In: CIKM'02. (2002)  60--67

\bibitem{chung2007erm}
Chung, Y.D., Yang, W.S., Kim, M.H.:
\newblock An efficient, robust method for processing of partial top-k/bottom-k
  queries using the {RD}-tree in {OLAP}.
\newblock Decision Support Systems \textbf{43}(2) (2007)  313--321

\bibitem{seokjin:eer}
Seokjin, H., Moon, B., Sukho, L.:
\newblock Efficient execution of range top-k queries in aggregate r-trees.
\newblock IEICE -- Transactions on Information and Systems \textbf{E88-D}(11)
  (2005)  2544--2554

\bibitem{donjerkovic1999pot}
Donjerkovic, D., Ramakrishnan, R.:
\newblock Probabilistic optimization of top n queries.
\newblock In: VLDB'99. (1999)  411--422

\bibitem{1232252}
Feige, U., Lee, J.R.:
\newblock An improved approximation ratio for the minimum linear arrangement
  problem.
\newblock Inf. Process. Lett. \textbf{101}(1) (2007)  26--29

\bibitem{bhasker1987ola}
Bhasker, J., Sahni, S.:
\newblock Optimal linear arrangement of circuit components.
\newblock J. VLSI Comp. Syst. \textbf{2}(1) (1987)  87--109

\bibitem{johnson2004clb}
Johnson, D., Krishnan, S., Chhugani, J., Kumar, S., Venkatasubramanian, S.:
\newblock Compressing large boolean matrices using reordering techniques.
\newblock In: VLDB'04. (2004)  13--23

\bibitem{pqtrees}
Booth, K.S., Lueker, G.S.:
\newblock Testing for the consecutive ones property, interval graphs, and graph
  planarity using {PQ}-tree algorithms.
\newblock Journal of Computer and System Sciences \textbf{13} (1976)  335--379

\bibitem{KDDRepository}
Hettich, S., Bay, S.D.:
\newblock The {UCI} {KDD} archive.
\newblock \url{http://kdd.ics.uci.edu}~(checked 2008-04-28) (2000)

\bibitem{253268}
O'Neil, P., Quass, D.:
\newblock Improved query performance with variant indexes.
\newblock In: SIGMOD '97. (1997)  38--49

\end{thebibliography}
\end{document}